# WEAK TEXTURE INFORMATION MAP GUIDED IMAGE SUPER-RESOLUTION WITH DEEP RESIDUAL NETWORKS

Bo Fu [1,2], Liyan Wang[1], Yuechu Wu[1], Yufeng Wu[1], Shilin Fu[1], Yonggong Ren[1]

1. School of Computer and Information Technology, Liaoning Normal University, China
2. School of SoftWare Technology, Dalian University of Technology

fubo@lnnu.edu.cn    ryg@lnnu.edu.cn

**Abstract.** Single image super-resolution (SISR) is an image processing task which obtains high-resolution (HR) image from a low-resolution (LR) image. Recently, due to the capability in feature extraction, a series of deep learning methods have brought important crucial improvement for SISR. However, we observe that no matter how deeper the networks are designed, they usually do not have good generalization ability, which leads to the fact that almost all of existing SR methods have poor performances on restoration of the weak texture details. To solve these problems, we propose a weak texture information map guided image super-resolution with deep residual networks. It contains three sub-networks, one main network which extracts the main features and fuses weak texture details, another two auxiliary networks extract the weak texture details fallen in the main network. Two part of networks work cooperatively, the auxiliary networks predict and integrates weak texture information into the main network, which is conducive to the main network learning more inconspicuous details. Experiments results demonstrate that our method's performs achieve the state-of-the-art quantitatively. Specifically, the image super-resolution results of our method own more weak texture details.

## 1  Introduction

Single image super-resolution (SISR) is wide range of applications in nature imaging, satellite imaging, medical imaging, security surveillance imaging and computer vision analysis task[1,2,3,37]. When various kinds of images or videos are usually caused by low-cost sensor, their quantities usually are certain restricted by clarity of lens, transmission bandwidth, sampling size, scene details, and other factors. Thus, search of super-resolution technology is a problem which restore super-resolution (SR) image from low-resolution (LR) image as similar as high-resolution (HR) image. However, solving super-resolution image is an ill-posed problem, since there exists multiple solutions for any LR image input. So far, a variety of super-resolution image reconstruction methods have emerged, which can be simply classified into three categories, i.e. linear interpolation function based, knowledge priors based, and example learning based methods.

The super-resolution methods based on linear interpolation function usually are simple and quite straightforward, such as Bilinear or Bicubic and S-Spline[4,5]. However, this kind of the method reconstruct by the local information, so usually work badly when a larger magnification factor is desired. The super-resolution methods based on knowledge priors relay on the representative priors such as the local structure similarity, some edge priors and other constraint information. Dai et al. used image patches as the local represent model, and reconstructed descriptors between background/foreground[6]. Sun et al. added the gradient prior information into the local image structure[7]. Glasner et al.[8] constructed a scale-space pyramid to exploit the self-similarity in given LR image. Compared with the first category of methods, the super-resolution methods based on knowledge priors is better at the expression of the detail texture. At the same time, this category of methods also can combine with other machine learning methods such as sparse-coding, Low-Rank Kernel, clustering and PCA [10,11,12,13,14]. But most of them are lack of a step which fix an invariant set of the parameters of represent models by learning. Therefore, the over-smoothing or blur phenomena often happen when these approaches process the natural images with different texture contents. The super-resolution methods example learning based methods attempt to reconstruct HR image from massive amount of LR-HR image dataset. Among these categories of methods, the deep neural network based methods[15,16,17,18,19,20,21,22] have achieved significant improvements.

Dong firstly introduced a three-layers CNN model to solve the SISR problems[9], named super-resolution convolutional neural network (SRCNN). SRCNN only owns three-layers networks architecture but get significantly outperforms than non-deep learning algorithms. Further, Kim et al. increased the network depth to 20 layers in named very deep convolutional networks (VDSR)[23] and deeply recursive convolutional network (DRCN)[24], which achieving the notable improvements over SRCNN. With the increase of neural network layers, the super resolution reconstruction methods based on deep learning also faces the problems of the gradient disappearance and the accuracy no improved. In order to solve the problem of the gradient disappearance, He at al.[25] proposed the residual net (ResNet). The residual learning strategy can effectively increase the layers of network, Lim et al.[26] built a very wide network by using simplified residual blocks. Tai et al.[27] used both the global residual connections and the local residual connections in deeply recursive residual networks (DRRN). The global residual learning is used in the identity branch and the recursive learning in the local residual branch. Mao et al.[28] proposed a 30-layer convolutional auto-encoder network, namely the residual encoder-decoder network (RED30). With the continuous development of deep learning theory, more CNN architecture such as DenseNet, Network in Network, and Residual Network have been exploited for SISR applications[29,30].

However, we find these networks mentioned above still exists some limitations in terms of network architectures. All those methods are designed to dig the notable features and restrain the unconspicuous features. Although this design can effectively reconstruct the most of details of image, it also discards some tiny details. To solve these problems, we propose a novel deep leaning framework which owns three sub-networks, one main network which extracts the main features and another two auxiliary networks which extract the weak texture details fallen in the main network. Three sub-networks work cooperatively, two auxiliary networks contribute more weak details into the main network, which is conducive to the main network learning more details.

The remainder of this paper is organized as follows: In section 2, we summarize and classify the existing SISR methods based on deep learning. In section 3, we introduce our network framework in detail. In Section 4, we show numerous experimental results and comparisons with several related algorithms on several types of nature images.

## 2 RELATED WORKS

For a SISR task, given a training set $\{I_{LR}, I_{HR}\}$, which contains several inputs low resolution images named $I_{LR}$ and its result image named $I_{SR}$. The goal of SISR task is make all $I_{SR}$ as similar as counterparts Ground-truth images named $I_{HR}$. So it is an essentially process of training the parameter set of networks and minimizing the loss function as follows.

$$\min L(F_{net}(I_{LR}), I_{HR}) + \lambda R \quad (1)$$

Here, $F_{net}$ denotes the function of the network, $\lambda R$ denotes the regularization term, and $L$ denotes the loss function of the network. As Formula 1 mentioned above, the first half part describes the neural network and the second half part describes the constraints of the network. Several witnessed significant SISR methods based on CNN have been studied in the computer vision community. These methods can be roughly divided into two categories as follow.

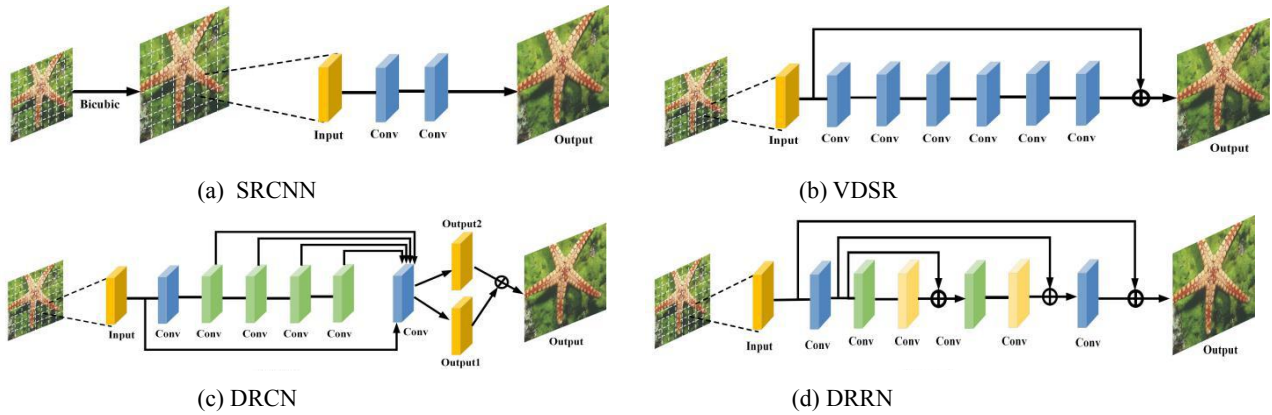

(a) SRCNN  (b) VDSR

(c) DRCN  (d) DRRN

Figure1. The diagram of several representative networks architecture

**SISR methods focusing on an advanced neural network architecture**. These kind of methods try to find more efficient network structure for SISR task. For example, Dong firstly introduced a three-layer CNN model to solve the SISR problems[9], and then Kim et al. increased the network depth to 20 layers in named very deep convolutional networks (VDSR)[23] and deeply recursive convolutional network (DRCN)[24]. It can be seen that these methods mentioned above are constantly revising the first half of Formula 1 formally. Among them, the Residual Channel Attention Network (RCAN) has achieved the-state-of-the-art (until 2019)[31]. The channel attention mechanism try to learn the spatial correlations between the filters operates with a local receptive field. In RCAN, the channel attention encapsulated as the Residual channel attention block modules(RCAB) and then several RCAB modules are introduced into the residual group(RG) ingeniously, its networks architecture are shown in Figure 2. Because of the effective combination of this attention mechanism and the residual network, RCAN can learn image features much better than the traditional CNN based methods.

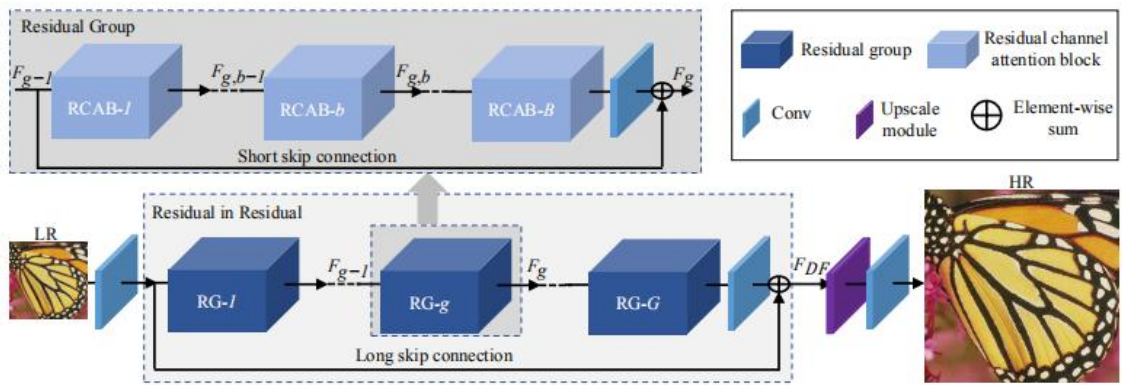

Figure 2 The diagram of the Residual Channel Attention Network (RCAN) architecture[31]

As shown in Figure 2, RCAN mainly consists four parts: the shallow feature extraction, residual in residual (RIR) deep feature extraction, the upscale module, and the reconstruction part.

**SISR methods focusing on increasing the priors information.** These kind of methods try to dig more efficient image priors information for SISR task. Yang et al. introduced self-similarities into the SISR task[34], and Jiang et al. defined two image priors constraint i.e. the non-local self-similarity and the local geometry priors[35]. Yang et al. used manifold localized as reference guided prior information and embed it into the networks[36], Fu et al. used TV model to generate the texture and enhanced in non-local frame which embed into the networks[19]. However, no matter the first kind of methods or the second kind of methods, they are often unsatisfactory in restoring the tiny image details. For the first type of SISR methods, they usually be good at extracting the most typical features but ignoring unconspicuous features. For the second type of SISR methods in our mention, they could not make sure the validity of the constraints though they added some priors information.

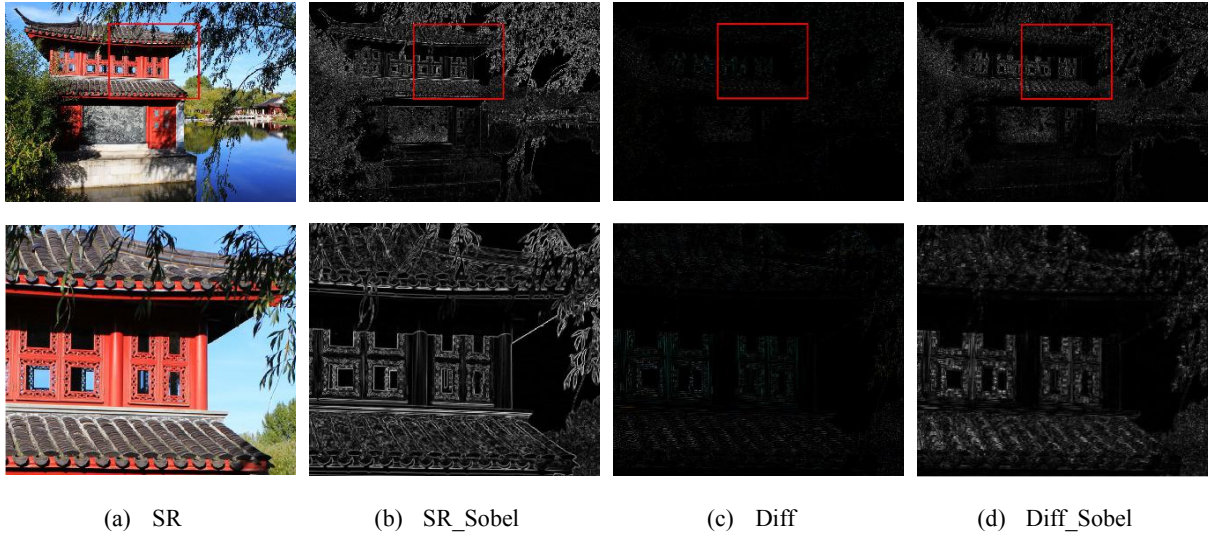

(a) SR      (b) SR_Sobel      (c) Diff      (d) Diff_Sobel

Figure 3    The diagram of artificial texture and real lost texture

For example, a generated texture and a true loss texture are shown in Figure 3(b) and 3(c), it can be seen that the lost texture information in networks usually is not conspicuous features such as the edges, the corners. The motivation of our work is to dig the real effective prior knowledge which is could not be learned from the mainstream SISR methods based on the deep learning.

## 3 Texture Information Map Guided Super-resolution with the Dual Residual Networks

In this section, we bridge a residual net with a pair of prediction networks for weak texture information. We describe the overall framework of our method firstly, then explain each module of our method and how to combine these modules effectively.

### 3.1 Network Architecture

As shown in Figure 4, our method mainly contains three sub-networks i.e. the RCAN model, the texture prediction model and the texture fusion model, respectively. In our network framework, the RCAN model is firstly adopt to generate a set of output graphs, named $I_{Output}$. $I_{Output}$ is used to generate a pair sets i.e. weak texture set named $I_{Output\_edge}$ and its corresponding differences set $I_{diff}$. The differences set $I_{diff}$ is the gray-value texture information which RCAN can't learn, and the weak texture information $I_{Output\_edge}$ is generated from $I_{Output}$. Then $I_{Output\_edge}$ will be input into the texture prediction model to learn the correlation between $I_{Output\_edge}$ and $I_{diff}$. As shown in Figure 5, the joint operation of the above two models is to predict the weak texture information set omitted by RCAN, named $I_{Edge}$. Because the last step generates additional dimensions of weak texture, those patches from set $I_{LR\_edge}$ and set $I_{LR}$ are integrated firstly and put into the texture fusion model. The essentially process of training parameter set of network and minimizing loss function as follows.

$$\min L(F_{TFM}(F_{TPM}(F_{RCAN}(I_{LR}))), I_{HR}) \tag{2}$$

Here, $F_{TFM}$, $F_{TPM}$, and $F_{RCAN}$ denote the function of the texture fusion model, the texture prediction model and the RCAN model.

### 3.2 Networks Description

Unlike classical Residual net like the RCAB [31], our framework adds a weak texture predication step. We denote $I_{LR}$ and $I_{SR}$ as the input and output data of our hole network framework. In training phase, $I_{LR}$ is input RCAN [31] model firstly, and its output result named $I_{Output}$.

$$I_{Output} = F_{RCAN}(I_{LR}) \tag{3}$$

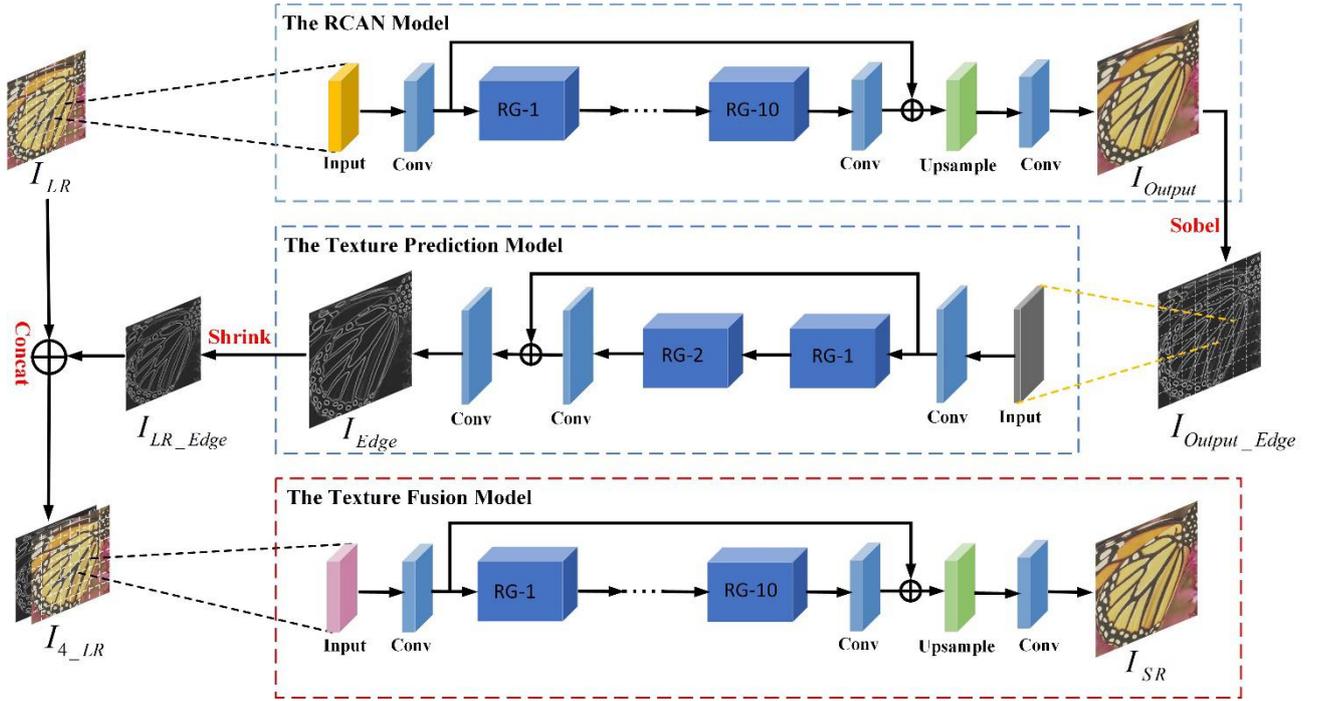

Figure 4 Network architecture of our method

Then Sobel operator is adopt to generate some dense edge information, named $I_{Output\text{-}Edge}$. In Figure 3, it can be noted that the second-order difference operator such as Sobel enhances either the information of strong texture and weak texture in same time. As shown in Figure 5, we construct a residual net named Texture Prediction Model(TPM) to learn the correlation between dense texture information set $I_{Output\text{-}Edge}$ and real texture information set $I_{Diff}$.

$$I_{Edge} = F_{TPM}(I_{Output\_Edge}) \tag{4}$$

In the Texture Prediction Model, we use one convolutional layer to extract the shallow feature from $I_{Output\text{-}Edge}$. Then data flow into the residual module which contains several residual groups to extract the deep feature. And then the data flow into a layer of convolution, its function is feature decoding. So formula 4 can further express in detail as follow.

$$I_{Edge} = F_{cov}(F_{sis}(F_{cov}(I_{Output\_Edge}))) \tag{5}$$

Here, $F_{sis}$ and $F_{cov}$ denote input residual groups operation and convolutional operation, respectively. The results $I_{Edge}$ obtained by the Texture Prediction Model, then it is shrink as according enlarge factor and named $I_{LR\_Edge}$. And *then $I_{LR\_Edge}$* is concatenated with original input data set $I_{LR}$ to generate a type of 4 dimensions data . We denote new confusion dataset as $I_{4\_LR}$. Then we conduct the Texture Fusion Model(TFM) as follows.

$$I_{SR} = F_{TFM}(I_{4\_LR}) \tag{6}$$

In the Texture Fusion Model, we also use one convolutional layer to extract the shallow feature , and then send these features into several residual groups, each residual group is consists of several residual blocks and short skip connection. More details about the residual group, the short skip connection and the attention mechanism have been demonstrated in [23,31]. A convolutional layer is adopt to reduce 4-dimensional data to 3 dimensions and another convolutional layer is used to up-sampling operation . The the formula 6 can be further expressed as follows.

$$I_{SR} = F_{reduce}F_{ups}(F_{sis}(F_{cov}(I_{4\_LR}))) \tag{7}$$

Here, $F_{reduce}$ and $F_{ups}$ denote the reduce dimension operation and the up-sampling operation, respectively.

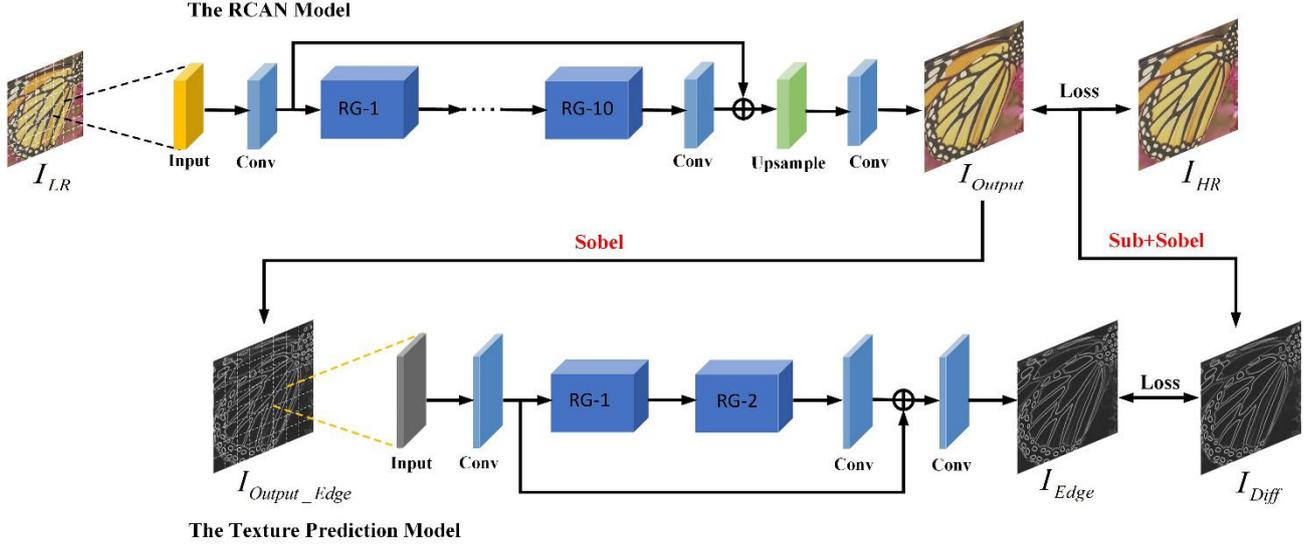

Figure 5 joint operation of the RCAN and TPM models to predict weak texture information

We construct a network framework which consist of three sub-models. Through the joint work of three sub-models, our network can dig and predict the weak texture and then fuse the weak texture information for new collaborative learning. More discussions about the effects of our method are shown in Section 4.2.

## 4 EXPERIMENTS

### 4.1 Experimental Setting

***Dataset*** We select DIV2K as training dataset, and some public datasets as experimental test dataset, such as the Set5, the Set14, B100, Manga109 and Urban100 datasets.

***Baseline algorithm*** We use six existing Super-Resolution algorithms as baselines, including Bicubic algorithm, FSRCNN[9], VDSR[23], MemNet[23], SRMDNF[17] and RCAN[31]. For fair comparison, all methods' code implementations are their publicly available versions and their parameters are set following the guidelines in original articles.

***Evaluation Metric*** The peak signal-to-noise ratio (PSNR) and the structural similarity index (SSIM) are adopted to measure the objective performance of our algorithm.

### 4.2 Network Parameters and Training Settings

Our network consists of 3 sub-models, and they are all have the shallow feature extractor, the residual groups. In the residual groups, there are several residual blocks, but the numbers of residual blocks in 3 sub-models are different. In RCAN model and the texture fusion model, we use 10 residual groups, each residual group owns 10 residual blocks. But in the texture prediction model we use only 2 residual groups, because the texture information of $I_{Output\text{-}Edge}$ is not dense enough. In each residual block, there are 2 convolutional layers with size of $3\times 3$ filters and 16 channel attention block modules with size of $H\times W\times C$ input data is squeezed into size of $1\times 1\times C$ data by averaging, then this $1\times 1\times C$ descriptors are put through the convolutional layers and gate activation of sigmoid function. During training step, the training pictures is cut into the patches with size of $48\times 48$. The batch size is set as 16, the learning rate is set as $10^{-4}$, and the ADAM is adopted as optimizer of 3 sub-models.

In our method, we only training 50 epoch for Texture Prediction Model, 200 epoch for Texture Fusion model. We choose same training dataset and iteration times for comparison methods such as RCAN and our method. All experiments are implemented on a workstation with NVIDIA Titan X GPU, and Core(TM) i7-7700K CPU.

### 4.3 Performance Comparison

Experiments were carried out on several common public test dataset. We trained RCAN again to make sure it is in same training dataset and training epoch environment with our method. Some other methods' sores can obtain by comparison their lectures. The PSNR scores and SSIM scores of our method and some state-of- the-art SISR methods are compared in Table 1.

Table 1 The PSNR and SSIM scores of our method and some state-of- the-art SISR methods

| Datasets | Evaluation Indexes | Bicubic | FSRCNN | VDSR | MemNet | SRMDNF | RCAN | Our method |
|---|---|---|---|---|---|---|---|---|
| Scale | | ×3 | ×3 | ×3 | ×3 | ×3 | ×3 | ×3 |
| Set5 | PSNR | 30.39 | 33.18 | 33.67 | 34.09 | 34.12 | 34.52 | **34.53** |
| | SSIM | 0.8682 | 0.9140 | 0.9210 | 0.9248 | 0.9254 | **0.9284** | 0.9282 |
| Set14 | PSNR | 27.55 | 29.37 | 29.78 | 30.00 | 30.04 | 30.34 | **30.46** |
| | SSIM | 0.7742 | 0.8240 | 0.8320 | 0.8350 | 0.8382 | 0.8432 | **0.8445** |
| B100 | PSNR | 27.21 | 28.53 | 28.83 | 28.96 | 28.97 | 29.11 | **29.17** |
| | SSIM | 0.7385 | 0.7910 | 0.7990 | 0.8001 | 0.8025 | 0.8063 | **0.8070** |
| Urban100 | PSNR | 24.46 | 26.43 | 27.14 | 27.56 | 27.57 | 28.32 | **28.56** |
| | SSIM | 0.7349 | 0.8080 | 0.8290 | 0.8376 | 0.8398 | 0.8577 | **0.8607** |
| Manga109 | PSNR | 26.95 | 31.10 | 32.01 | 32.51 | 33.00 | 33.54 | **33.80** |
| | SSIM | 0.8556 | 0.9210 | 0.9340 | 0.9369 | 0.9403 | 0.9449 | **0.9464** |

It can be seen that our algorithm obtain superiority in most test dataset either the PSNR or the SSIM index. At the same time, visual comparisons on enlarging three times of the image Butterfly are also provided in Figure 6, and the visual comparisons on enlarging three times of details enlarged are provided in Figure 7.

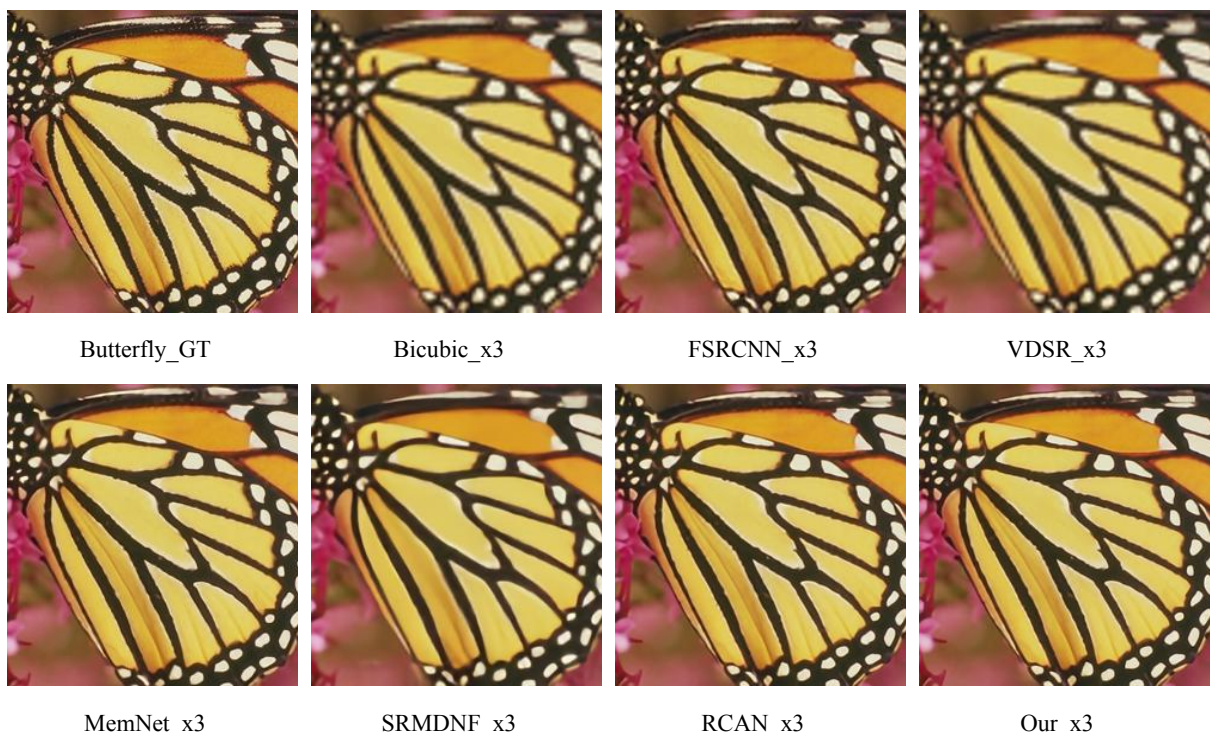

Butterfly_GT     Bicubic_x3     FSRCNN_x3     VDSR_x3

MemNet_x3     SRMDNF_x3     RCAN_x3     Our_x3

Figure 6 The visual quality results of the image Butterfly

It can be seen that our algorithm obtains good visual effect. Due to predict more weak texture information, our algorithm can get more tiny edge information better.

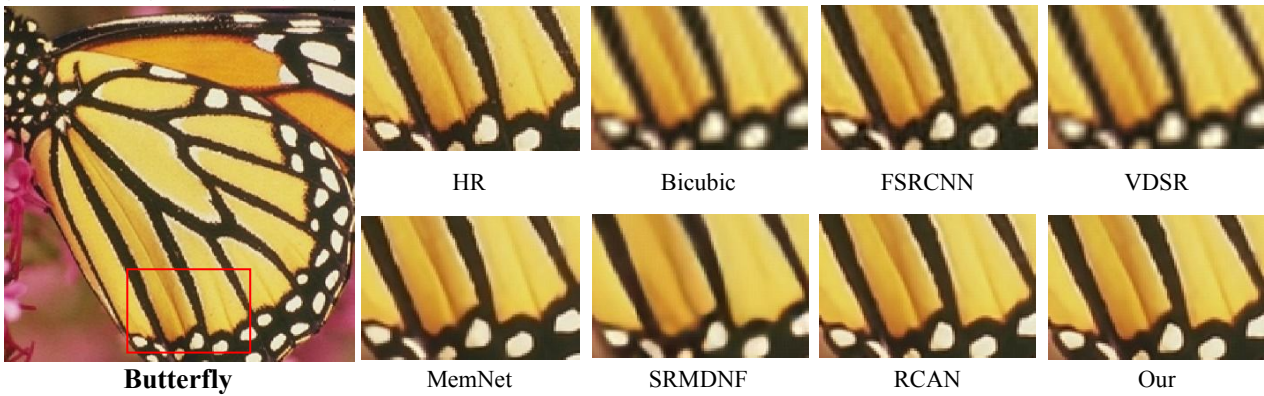

Figure 7    The visual quality results of enlarged the image Butterfly

## 5  CONCLUSION

In this paper, we propose a texture information map guided residual networks. It contains the main network and two auxiliary prediction networks. The main network extracts and combines the mainly obvious deep features and the auxiliary prediction networks extract the weak texture information. Three parts of the sub networks model work cooperatively, the two auxiliary networks predict and integrate more weak texture information into the main network, which is conducive to the main network learning more inconspicuous details. Experiments results demonstrate that our method's performs achieve the state-of-the-art quantitatively.

**ACKNOWLEDGEMENTS**

This work is supported by the National Natural Science Foundation of China (NSFC) Grant No.61702246, No.61976109, China Postdoctoral Science Foundation, No. 2019M651123 and Science and Technology Innovation Fund(Youth Science and Technology Star) of Dalian, China, No. 2018RQ65. Liaoning Natural Science Foundation (No.20180550542), Dalian Science and Technology Innovation Fund (No.2018J12GX047), Dalian Key Laboratory Special Fund.